\documentclass[11pt]{revtex4-1}
\usepackage[latin9]{inputenc}
\usepackage{geometry}
\usepackage{array}
\usepackage{float}
\usepackage{textcomp}
\usepackage{amsbsy}
\usepackage{amssymb}
\usepackage{amsmath}
\usepackage{graphicx}
\usepackage{esint}

\makeatletter

\providecommand{\tabularnewline}{\\}

\@ifundefined{textcolor}{}
{%
 \definecolor{BLACK}{gray}{0}
 \definecolor{WHITE}{gray}{1}
 \definecolor{RED}{rgb}{1,0,0}
 \definecolor{GREEN}{rgb}{0,1,0}
 \definecolor{BLUE}{rgb}{0,0,1}
 \definecolor{CYAN}{cmyk}{1,0,0,0}
 \definecolor{MAGENTA}{cmyk}{0,1,0,0}
 \definecolor{YELLOW}{cmyk}{0,0,1,0}
}

\usepackage[font=small]{caption}
\usepackage{epstopdf}
\usepackage{dcolumn}
\usepackage{bm}
\usepackage{array}
\usepackage{tabto}

\usepackage{color}

\makeatother

\begin{document}

\title{Mapping an electron wave function by a local electron scattering
probe}

\author{C. Reichl$^{1}$,}

\email{creichl@phys.ethz.ch}

\author{W. Dietsche$^{1,2}$, T. Tschirky$^{1}$, T. Hyart$^{3}$ and W.
Wegscheider$^{1}$}

\affiliation{$^{1}$Solid State Physics Laboratory, ETH Zurich, Otto-Stern-Weg
1, 8093 Zürich, Switzerland}

\affiliation{$^{2}$Max Planck Institut für Festkörperforschung, Heisenbergstrasse
1, 70569 Stuttgart, Germany}

\affiliation{$^{3}$University of Jyväskylä, Department of Physics and Nanoscience
Center, P.O. Box 35 (YFL), FI-40014, Finland}
\begin{abstract}
A technique is developed which allows for the detailed mapping of
the electronic wave function in two-dimensional electron gases with
low-temperature mobilities up to $15\cdot10^{6}\,\mathrm{cm^{2}/Vs}$.
Thin (``delta'') layers of aluminium are placed into the regions
where the electrons reside. This causes electron scattering which
depends very locally on the amplitude of the electron wave function
at the position of the Al $\delta$-layer. By changing the distance
of this layer from the interface we map the shape of the wave function
perpendicular to the interface. Despite having a profound effect on
the electron mobiliy, the $\delta$-layers do not cause a widening
of the quantum Hall plateaus. 
\end{abstract}
\maketitle

\section{Introduction}

The envelope wave function $\Psi(\boldsymbol{r})$ of localized electrons
in semiconductors is determined by the laws of quantum-mechanics and
electrostatics. Although the shape of $\Psi(\boldsymbol{r})$ determines
many physical properties, its precise form is experimentally only
accessible under very favourable conditions and with substantial effort,
for example using an UHV-STM \cite{2007Suzuki}. In this work we utilize
the extremely short interaction length of neutral impurities in high
quality GaAs/AlGaAs heterostructures, synthesized by molecular beam
epitaxy (MBE) to map out the square of the electron wave function
perpendicular to the interface. This requires to place very thin (``delta'')
layers of Al atoms at varying positions and measure the electron mobilities,
from which the electron scattering rates are determined. These scattering
rates reflect the amplitude of the wave function at the position of
the $\delta$-layer.

Electrons in two dimensional electron gases (2DEG) in heterostructures
are free to propagate along the interface but are localized perpendicular
to it\cite{IhnBook}. The eigenstates and eigenenergies in the absence
of a scattering potential are

\begin{equation}
\psi_{\mathbf{k}}(\mathbf{r},z)=\frac{1}{\sqrt{L^{2}}}e^{i\mathbf{k}\cdot\mathbf{r}}\chi(z),\ E(\mathbf{k})=E_{0}+\frac{\hbar^{2}k^{2}}{2m^{*}},
\end{equation}
where $\mathbf{r}=(x,y)$, $\mathbf{k}=(k_{x},k_{y})$, $\chi(z)$
is the normalized wave function for the lowest energy transverse mode.
The factor $\sqrt{L^{2}}$ is a normalization, $E_{0}$ is the ground
state eigenenergy and $m^{*}=0.067m_{e}$ is the effective mass.

The function $\chi(z)$ can be calculated self-consistently by combining
the Schrödinger and the Poisson equation. This requires assumptions
about the material parameters of the semiconductor structures, particularly
the boundary conditions, band offsets at the interface and the incorporation
of doping atoms. Several software packages are available for numerical
solutions \cite{Nextnano,1999Rother,Snyder}, which however suffer
for example from the lack of precise values for the band offsets\cite{2010Gerhardts}.
Consequently, while the theoretical model describing the wave function
is well established, $\chi(z)$ is typically obtained only approximately
by means of simulation.

Neutral impurities, e.g. atoms like Al with the same outer electron
shell as Ga, are known to have very short interaction lengths \cite{1984Walukiewicz,2003Li},
although details of the scattering mechanism have not yet been resolved.
Adding $\delta$-layers of Al to the GaAs in the region where the
2DEG resides should allow to test the amplitude of $\chi(z)$ at the
$\delta$-layer position. This is done by analyzing the electron mobility
$\mu$.

To utilize the Al $\delta$-layer as local scattering center, it is
of paramount importance to reduce all other scattering processes as
much as possible. These processes include scattering by charged ionized
donors, phonons, interface roughness and background impurities (see
e.g. \cite{1984Walukiewicz}). The latter stem from residuals in the
MBE chamber that are inevitably incorporated during the growth process.

The phonon scattering can be effectively removed by cooling the sample
to low temperatures. The role of the ionized donors is minimized by
large setback distances between doping and 2DEG, and the effect of
interface roughness appears to be negligible under optimized growth
conditions. The background impurities can only be reduced if the heterostructures
are synthesized under extreme purity in specialized molecular beam
epitaxy (MBE) setups. The ``quality'' of a given MBE setup is generally
measured by the maximum electron mobility which has been achieved
in quantum-well structures\cite{2002Arthur,2003Pfeiffer,2009Umansky,2010Schlom,2014Manfra}.
Mobilities exceeding $2.5\cdot10^{7}\,\mathrm{cm^{2}/Vs}$ (measured
at $300\,\mathrm{mK}$) have been achieved with the MBE setup used
by us for growing the Al-doped samples\cite{2015DisIch}. 
\begin{figure}[H]
\begin{centering}
\includegraphics[width=0.6\textwidth]{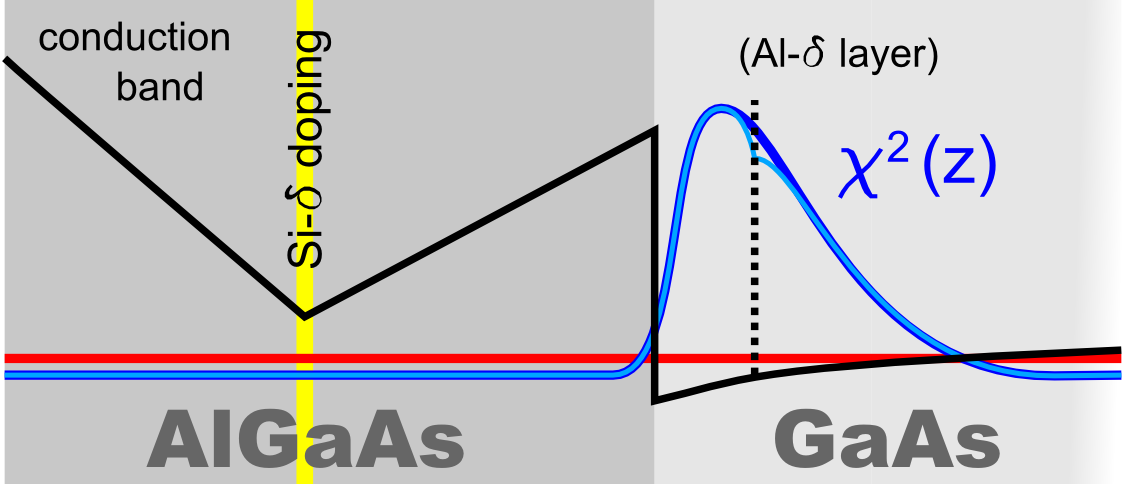} 
\par\end{centering}

\protect\protect\caption{Schematic of the heterostructures. The black line illustrates the
conduction band along the growth direction (with the sample surface
towards the left), the function $\chi^{2}(z)$ is the squared envelope
wave function. The grey area denotes the AlGaAs region, marked in
yellow is the Si doping layer. The black dotted line illustrates the
band shape with an included layer of neutral impurities (aluminium),
leading to a small deviation in the shape of the wave function (light
blue line). \textbf{\label{MDSI}}}
\end{figure}

\section{Experimental details}

As the basic sample design we use single-sidedly doped heterostructures
(Fig. \ref{MDSI}) grown in the following sequence: We start with
a superlattice consisting of $100$ periods of $7\,\mathrm{nm}$ AlGaAs
and $3\,\mathrm{nm}$ GaAs. This is followed by $1000\,\mathrm{nm}$
GaAs hosting the 2DEG at the interface to an adjacent $310\,\mathrm{nm}$
thick AlGaAs layer. This region contains a thin doping layer of silicon,
placed at a setback distance of $70\,\mathrm{nm}$ from the interface.
The whole structure is capped by $10\,\mathrm{nm}$ of GaAs. The Al
content of the AlGaAs is $25\,\%$ throughout. These values are based
on a growth rate calibration that is performed on a daily basis and
secures that rates and with that layer thicknesses are precise within
a margin of less than $2\,\%$ (see \cite{2015DisIch} for a detailled
description). A series of different samples are grown containing Al
impurities which replace the Ga atoms in the GaAs crystal structure.
This is done by adding $0.28\,\mathrm{nm}$ (one monolayer) AlGaAs
and with that $1.5\cdot10^{14}\,\mathrm{cm^{-2}}$ Al atoms to the
GaAs at distances $a$ from the interface varying from $a=5\,\mathrm{nm}$
to $a=30\,\mathrm{nm}$. The average distance between the Al atoms
in this layer is $0.8\,\mathrm{nm}$. The dispersion of the AlGaAs
delta layer due to migration during the growth process can be considered
negligible. TEM analysis of structures produced under similar growth
conditions shows sharp interfaces of an AlAs layers of $2\,\mathrm{nm}$
width. The same is true for a buffer superlattice as described above.
A TEM image of such a superlattice with comparable interface quality
is shown in \cite{2014Manfra}.

Transport properties were measured by the van-der-Pauw technique,
both in the dark and after illumination with a red ($710\,\mathrm{nm}$)
LED. Magnetotransport data were obtained at $1.3\,\mathrm{K}$ at
magnetic fields up to $6$ Tesla. The electron densities and mobilities
at $1.3\,\mathrm{K}$ of the sample without any Al impurities are
$1.5\cdot10^{11}\,\mathrm{cm^{-2}}$ and $8.0\cdot10^{6}\,\mathrm{cm^{2}/Vs}$
in the dark and $2.0\cdot10^{11}\,\mathrm{cm^{-2}}$ and $14\cdot10^{6}\,\mathrm{cm^{2}/Vs}$
after illumination, respectively. This structure serves as the reference
for the series with Al $\delta$-layers at varying distances and will
furtheron be referred to as $a=0$.

\section{The scattering vs. Al-doping depth}

\begin{table}
\begin{centering}
\begin{tabular}{>{\centering}p{1.5cm}|>{\centering}p{2cm}>{\centering}p{2cm}|>{\centering}p{2cm}}
distance {[}$\mathrm{nm}${]}  & density {[}$10^{11}\,\mathrm{cm^{-2}}${]}  & mobility {[}$10^{6}\,\mathrm{cm^{2}/Vs}${]}  & scattering rate {[}$\mathrm{ns^{-1}}${]}\tabularnewline
\hline 
\hline 
0  & 1.52  & 8.036  & 3.27 \tabularnewline
5  & 1.40  & 1.293  & 20.30 \tabularnewline
10  & 1.39  & 0.793  & 33.1 \tabularnewline
15  & 1.416  & 1.813  & 14.48\tabularnewline
20  & 1.461  & 3.746  & 7.01 \tabularnewline
25  & 1.561  & 7.331  & 3.58 \tabularnewline
30  & 1.574  & 8.281  & 3.17 \tabularnewline
\end{tabular}
\par\end{centering}

\protect\protect\caption{Electron densities and mobilities of the samples with different distances
of the Al $\delta$-layers from the interface. Also shown are the
scattering rates calculated from the mobilities. All data are obtained
in the dark. \textbf{\label{dark data}}}
\end{table}

\begin{table}
\begin{centering}
\begin{tabular}{>{\centering}p{1.5cm}|>{\centering}p{2cm}>{\centering}p{2cm}|>{\centering}p{2cm}}
distance {[}$\mathrm{nm}${]}  & density {[}$10^{11}\,\mathrm{cm^{-2}}${]}  & mobility {[}$10^{6}\,\mathrm{cm^{2}/Vs}${]}  & scattering rate {[}$\mathrm{ns^{-1}}${]} \tabularnewline
\hline 
\hline 
0  & 1.98  & 13.75  & 1.91\tabularnewline
5  & 1.79  & 1.98  & 13.255 \tabularnewline
10  & 1.76  & 1.14  & 22.98 \tabularnewline
15  & 1.81  & 2.309  & 11.37 \tabularnewline
20  & 1.96  & 5.21  & 5.04 \tabularnewline
25  & 1.99  & 9.53  & 2.75 \tabularnewline
30  & 2.04  & 12.77  & 2.05 \tabularnewline
\end{tabular}
\par\end{centering}

\protect\protect\caption{Characterization data as in table \ref{dark data} but after illumination
with a red LED. \textbf{\label{LED data}}}
\end{table}

The tables \ref{dark data} and \ref{LED data} summarize the results
for measurements made in the dark and after illumination at $1.3\,\mathrm{K}$,
respectively. As usual, the illuminated structures have larger electron
densities compared to those measured in the dark. Adding the Al $\delta$-layers
has a significant effect on electron mobility: $\mu$ drops by an
order of magnitude from $14\cdot10^{6}\,\mathrm{cm^{2}/Vs}$ to $1.1\cdot10^{6}\,\mathrm{cm^{2}/Vs}$
after illumination, and from $8.0\cdot10^{6}\,\mathrm{cm^{2}/Vs}$
to $0.8\cdot10^{6}\,\mathrm{cm^{2}/Vs}$ in the dark), if the Al is
placed $10\,\mathrm{nm}$ away from the interface (which is the mobility
range of what Gardner et al. reported for a comparable, homogeneously
distributed amount of Al atoms \cite{2013Gardner}). We note that
we are able to reproduce the magnetotransport characteristics of nominally
identical samples, originating from different growth runs within a
margin of $2\,\%$ for electron density and $4\,\%$ for mobility\cite{2014ich}.

It is useful to compare the transport scattering rates $1/\tau$ rather
than the mobilities to discriminate the intrinsic scattering processes
-- caused by background impurities in the growth chamber, remote ionized
donor potential disorder, interface roughness and phonon scattering
-- from the ones induced exclusively by the Al impurities. The total
scattering rate $1/\tau_{tot}$ should be the sum of the intrinsic
rate $1/\tau_{int}$ and the one due to the Al impurities $1/\tau_{Al}$:

\begin{equation}
1/\tau_{tot}=1/\tau_{int}+1/\tau_{Al}
\end{equation}

$1/\tau_{tot}$ is calculated from the relation $\mu=(e/m^{\ast})\cdot\tau_{tot}$,
where $e$ and $m^{\ast}$ are the elementary charge and the effective
electron mass, respectively.

\begin{figure}
\begin{centering}
\includegraphics[width=0.5\textwidth]{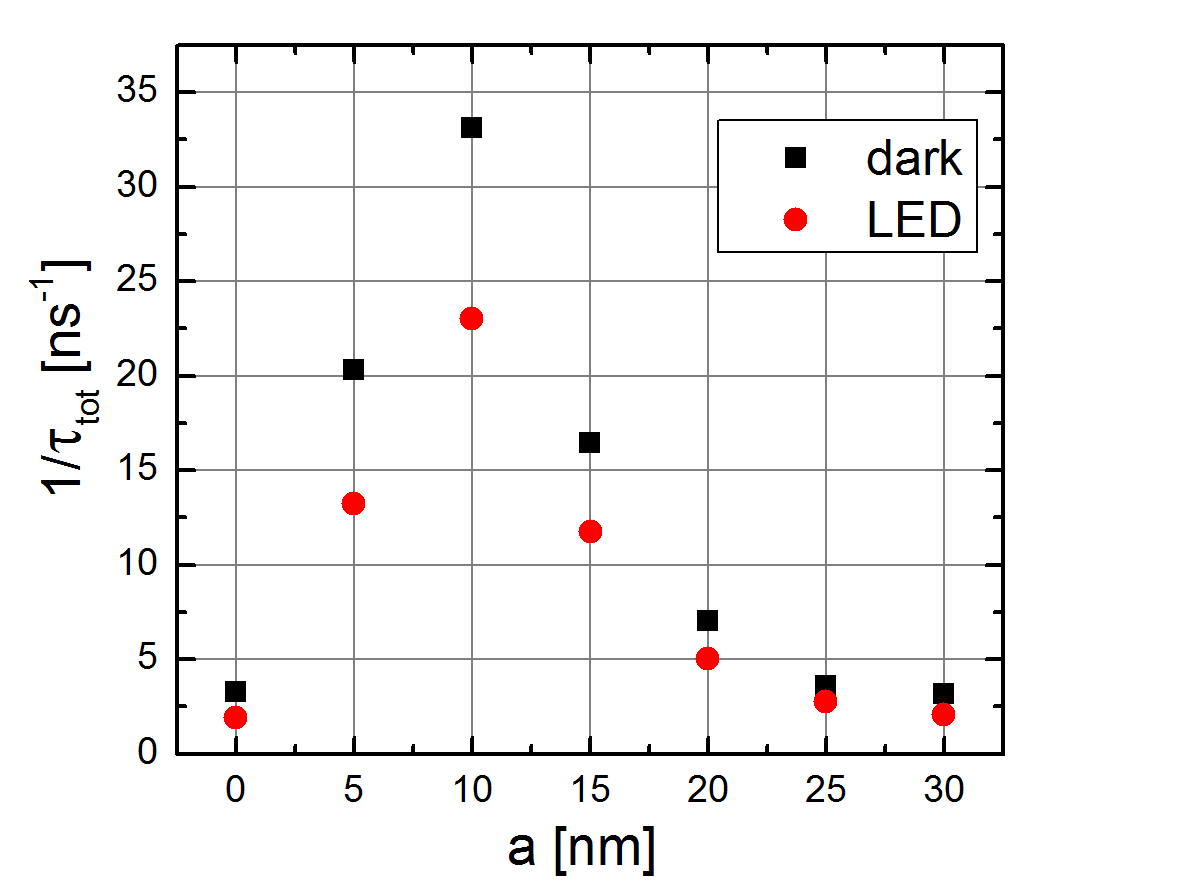}
\par\end{centering}

\protect\protect\caption{Scattering rates $1/\tau_{tot}$ as a function of the distance $a$
of the Al $\delta$-layer from the interface. The black squares represent
data measured in the dark, the red circles are obtained after illumination.\textbf{\label{ScatterRatesOnly}} }
\end{figure}

The resulting scattering rates are shown in tables \ref{dark data}
and \ref{LED data} and are plotted in Fig. \ref{ScatterRatesOnly}.
Unexpectedly, the electron density is reduced by up to $10\,\%$,
if the Al $\delta$-layer is located in the $5$ to $15\,\mathrm{nm}$
range. This systematic change is too large to be accustomed to uncertainties
in the growth process (those may account for a density variation of
no more than $1\,\%$) or the error margin of the characterization.

The scattering rates have a maximum at a distance of $10\,\mathrm{nm}$
from the interface where they exceed the reference values by a factor
of about $10$. It is noteworthy that not only the reference scattering
rate but also the one due to the Al atoms decrease after illumination.

\section{Scattering by neutral impurities -- Theory\label{sec:Scattering-by-neutral}}

Although a first principle calculation of the scattering is difficult,
a simple approximation can be obtained by modifying the approach used
in \cite{1976Harrison,1983Basu} in such a way that the scattering
sites now exist only in the Al $\delta$-layer. In this approach one
considers the Ga atoms being replaced by Al atoms randomly in some
sites $\mathbf{r}_{i}$ (possibly also clustering around these sites,
such that $\mathbf{r}_{i}$ are the centers of these clusters). The
average ($V_{{\rm av}}(\mathbf{r},z)$) and the random ($V_{{\rm rand}}(\mathbf{r},z)$)
part of the potential are 
\[
V_{{\rm av}}(\mathbf{r},z)=\delta(z-a)W\sum_{i}\big[(1-x)V_{{\rm Ga}}(\mathbf{r}-\mathbf{r}_{i})+xV_{{\rm Al}}(\mathbf{r}-\mathbf{r}_{i})\big]
\]

\begin{equation}
V_{{\rm rand}}(\mathbf{r},z)=\delta(z-a)W\sum_{i}C_{i}\big[V_{{\rm Ga}}(\mathbf{r}-\mathbf{r}_{i})-V_{{\rm Al}}(\mathbf{r}-\mathbf{r}_{i})\big],
\end{equation}
where $W=0.28\,\mathrm{nm}$ and $a$ are the approximate thickness
and the position of the Al $\delta$-layer, respectively, and each
$C_{i}$ is a random variable which is $x$ with a probability of
$1-x$ and $x-1$ with probability $x$. Here $x=0.25$ is the Al
concentration in the $\delta$-layer. We assume that $C_{i}$ in different
sites are uncorrelated so that the expectation value over the disorder
realizations satisfies $\langle C_{i}C_{j}\rangle=x(1-x)\delta_{ij}$.

The homogeneous average potential $V_{{\rm av}}(\mathbf{r})$ does
not cause scattering, so that the scattering rate is completely determined
by the random potential $V_{{\rm rand}}(\mathbf{r})$. We parametrize
the potential around each site $\mathbf{r}_{i}$ as 
\begin{equation}
V_{{\rm Ga}}(\mathbf{r}-\mathbf{r}_{i})-V_{{\rm Al}}(\mathbf{r}-\mathbf{r}_{i})=\Delta E\ H(r_{0}-|\mathbf{r}-\mathbf{r}_{i}|),
\end{equation}
where $H(x)$ is the Heaviside step function, and $\Delta E$ and
$r_{0}$ describe the magnitude and the range of the scattering potential
caused by each cluster. The scattering rate can be calculated using
Fermi's golden rule 
\begin{equation}
\frac{1}{\tau(k)}=\frac{2\pi}{\hbar}\frac{L^{2}}{(2\pi)^{2}}\int dk'\int d\theta\ k'\ \langle|M(\mathbf{k},\mathbf{k}')|^{2}\rangle\delta\big(E(\mathbf{k})-E(\mathbf{k}')\big)(1-\cos\theta),
\end{equation}
where $\theta$ is the angle between $\mathbf{k}$ and $\mathbf{k}'$
and $M(\mathbf{k},\mathbf{k}')=\int d^{2}r\int dz\ \psi_{\mathbf{k}}^{*}(\mathbf{r},z)V_{{\rm rand}}(\mathbf{r},z)\psi_{\mathbf{k}'}(\mathbf{r},z)$
is the matrix element caused by the random alloy scattering potential.
By assuming that the Fermi wavevector $k_{F}$ satisfies $k_{F}r_{0}\ll1$,
we obtain 
\begin{equation}
\frac{1}{\tau}=\pi^{2}\chi(a)^{4}W^{2}\frac{(\Delta E)^{2}r_{0}^{4}m^{*}}{\hbar^{3}L^{2}}x(1-x)N,\label{eq:scattering rate}
\end{equation}
where $N$ is the number of scattering sites.

This formula shows that the scattering rate is proportional to the
fourth power of the wave function at $z=a$. This behavior is used
for the wave function mapping. The dependence of the scattering rate
on the fourth power of $\chi(z)$ leads to the rapid variation of
the scattering rates with the distance from the interface as seen
in the data presented in Fig. \ref{ScatterRatesOnly}.

For a quantitative estimate of the scattering rate one has to make
assumptions about $r_{0}$ and the the scattering potential $\Delta E$.
We assume that the range $r_{0}$ and the spacing between the scattering
sites $\sqrt{L^{2}/N}$ are on the same order $r_{0}\approx\sqrt{L^{2}/N}\approx1\,{\rm nm}$.
Then the only free parameter is the magnitude of the scattering potential
$\Delta E$, which is expected to be on the order of $\Delta E\sim0.1-1$
eV corresponding to the conduction band variations if Al atoms are
alloyed to the GaAs. With $\chi^{2}(a)\approx0.06\,({\rm nm})^{-1}$
at $a=10\,{\rm nm}$ (see Fig.\ref{DataAndWave}) and a $\Delta E\approx0.2\,\mathrm{eV}$
we find a scattering rate $\tau^{-1}\approx28\,({\rm ns})^{-1}$ which
is very close to the numbers measured.

Although the value of $\Delta E$ deduced from this analysis is considerable
smaller than the one found by Li et al. \cite{2003Li} for GaAs homogeneously
doped with Al, we believe that this discrepancy arises mainly from
model-specific assumptions that influence the value obtained for $\Delta E$.
In particular the parameters $r_{0}$ and $L^{2}/N$ have significant
uncertainties due to the possible clustering of the atoms, and in
this work we have made different assumptions for these parameters
than in \cite{2003Li}. Despite these uncertainties in the relative
magnitudes of the parameters, this theoretical calculation illustrates
that the experimentally measured scattering rates are consistent with
the alloy scattering mechanisms since quantitative agreement can be
obtained with a reasonable choice of the parameters $\Delta E$, $r_{0}$
and $L^{2}/N$.

\section{Determining the wave-function shape}

\begin{figure}
(a)\includegraphics[width=0.6\textwidth]{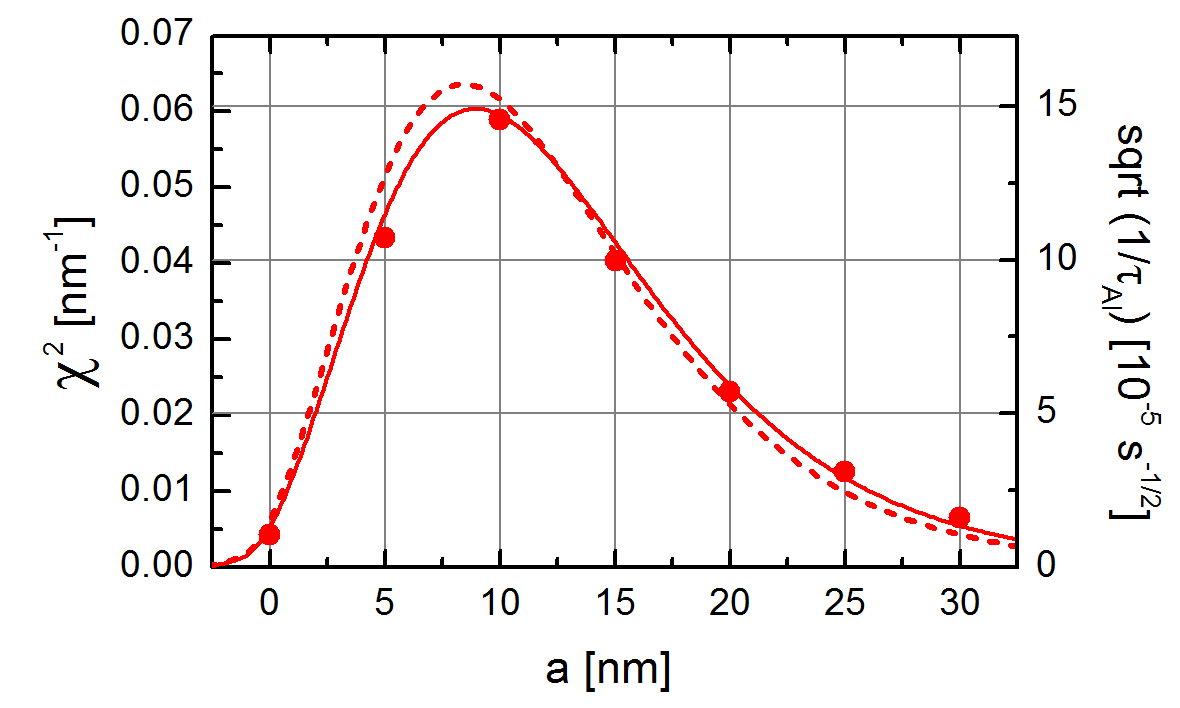}\\

(b)\includegraphics[width=0.6\textwidth]{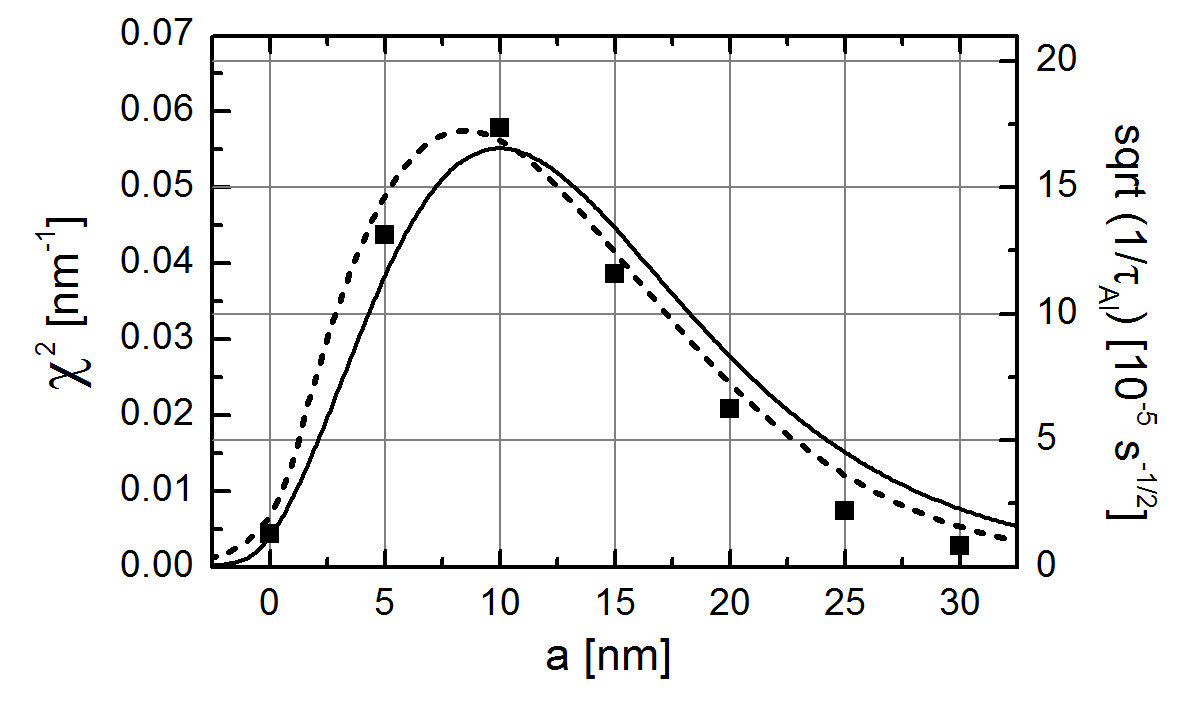}

\protect\protect\caption{Square root of the scattering rates $1/\tau_{Al}$ (dots) as a function
of the distance of the Al $\delta$-layer from the interface. Data
obtained after illumination and in the dark are shown in (a) and (b),
respectively. The solid lines correspond to the respective $\chi^{2}(z)$
as obtained from the Schrödinger-Poisson solver with adapted Si doping
density. The agreement between experimental data and theoretical curve
is very good, especially for the illuminated case. The dashed line
in (a) is the result of a calculation where the actual Si doping density
was used. In (b) the dashed line represents simulation data that includes
an impurity background in the initial AlGaAs layers at the beginning
of the growth process.\textbf{\label{DataAndWave}}}
\end{figure}

Based upon the data presented in Fig.\ref{ScatterRatesOnly}, we use
equation \ref{eq:scattering rate} to deduce the shape of the squared
envelope wave function $\chi^{2}(a)$. First, one needs to substract
an estimate of $1/\tau_{int}$ ($3.1\,\mathrm{ns^{-1}}$ and $1.8\,\mathrm{ns^{-1}}$
for the dark and illuminated state, respectively). The square root
of the resulting $1/\tau_{Al}$ is plotted as dots in Fig.~\ref{DataAndWave}(a)
and (b) for the illuminated and the non-illuminated case respectively.

These data points can be compared with theoretically expected wave
functions $\chi^{2}(a)$%
\footnote{For this comparison between measurement and simulation data we use
as a first order approximation wave function as obtained by simulation
without the Al $\delta$-layer. This appears to be a reasonable approximation,
since including the Al layer into the simulation leads to a density
variation of less than $0.5\,\%$ and a maximum change in $\chi^{2}(a)$
of less than $10\,\%$. %
} of the 2DEGs, obtained from the 8-band Schrödinger-Poisson-solver
software Nextnano\cite{Nextnano} which uses parameters from \cite{2001Vurgaftman},
including a conduction band offset of $250\,\mathrm{meV}$ for an
Al-fraction of $25\,\%$. The simulated structure is identical to
the actual samples, including a silicon doping layer with a density
of $3\cdot10^{12}\,\mathrm{cm^{-2}}$. Since the simulation neglects
the formation (and fraction) of DX-centers, the resulting wave function
is only applicable to the illuminated case, when almost all DX-centers
are ionized. The resulting wave function is shown as the dashed red
curve in Fig.~\ref{DataAndWave}(a), its calculated electron density
is higher ($2.25\cdot10^{11}\,\mathrm{cm^{-2}}$) than what was observed
experimentally ($\thicksim1.9\cdot10^{11}\,\mathrm{cm^{-2}}$); however,
its agreement with the experimental data is already very good and
gives trust in the mapping technique used here.

The fit can even be improved by adjusting the density of active donors
in the simulation to find a 2DEG density that matches the measured
one. This approach leads to the red solid line in Fig.~\ref{DataAndWave}(a),
which agrees excellently with the data points.

Fig.~\ref{DataAndWave}(b) plots the data obtained in the dark. Using
the the procedure as in the illuminated case, including an adjustment
in the active donor density (represented by the black solid line),
leads to a less good agreement with the data, particularly on the
wave function's flank far from the interface. This hints that the
2DEG is more strongly confined than anticipated by the simulation
software. Such an enhanced confinement could be the result of deep
level p-type impurites gettered by the highly reactive aluminium in
the AlGaAs/GaAs superlattice located far below the actual heterostructure.
The dashed line in Fig.~\ref{DataAndWave}(b) exemplarily shows the
resulting wave function for a background impurity density of $10^{15}\,\mathrm{cm^{-3}}$
in the AlGaAs buffer layers%
\footnote{Note that a background impurity density this high is assumed only
for the inital stages of the growth run and might for example stem
from the oxide layer protecting the substrate surface before growth.
We further assume that the background impurity level is not constant
during the growth run but is continuously reduced by gettering/pumping.
A mobility-density analysis as described in \cite{2010Mak} was performed
by the Ritchie group on a comparable structure grown by us and suggests
a charged background impurity of $\approx4\cdot10^{13}\,\mathrm{cm^{-3}}$
in the 2DEG region. %
}. Using this scenario, the experimentally observed wave function can
be reproduced very well for the dark case also. By means of illumination,
the background impurities in the buffer layers might be compensated,
leading back to the situation described above for the illuminated
case.

Overall, the agreement of the fit and the experimental data is surprisingly
good, from which we conclude that the scattering potential of the
Al atoms acts very locally on the electron wave function. It is noteworthy
that not only the intrinsic scattering rates but also $1/\tau_{Al}$
are reduced after illumination. The intrinsic scattering is probably
due to charged impurities, both from the Si-doping and in the 2DEG
region. In both cases screening has always been considered to be very
effective. Our data indicate that for the scattering by neutral impurities,
a density dependence exists, which also cannot be explained by the
shift of the wave function due to the illumination. Such a dependency
has, however, been neglected in previous theories\cite{1984Walukiewicz}
and is also not part of our analysis in section \ref{sec:Scattering-by-neutral}.

\section{The effect on magnetotransport\label{sec:Magnetotransport}}

In high perpendicular magnetic fields, the electronic transport properties
show the integer quantum Hall effect (IQHE). Generally, the widths
of the plateaus and the accompanying minima in the resistance depend
on the density of localized states between the Landau levels \cite{1981AOKI}
containing the extended states.

\begin{figure}
\begin{centering}
\includegraphics[width=0.8\textwidth]{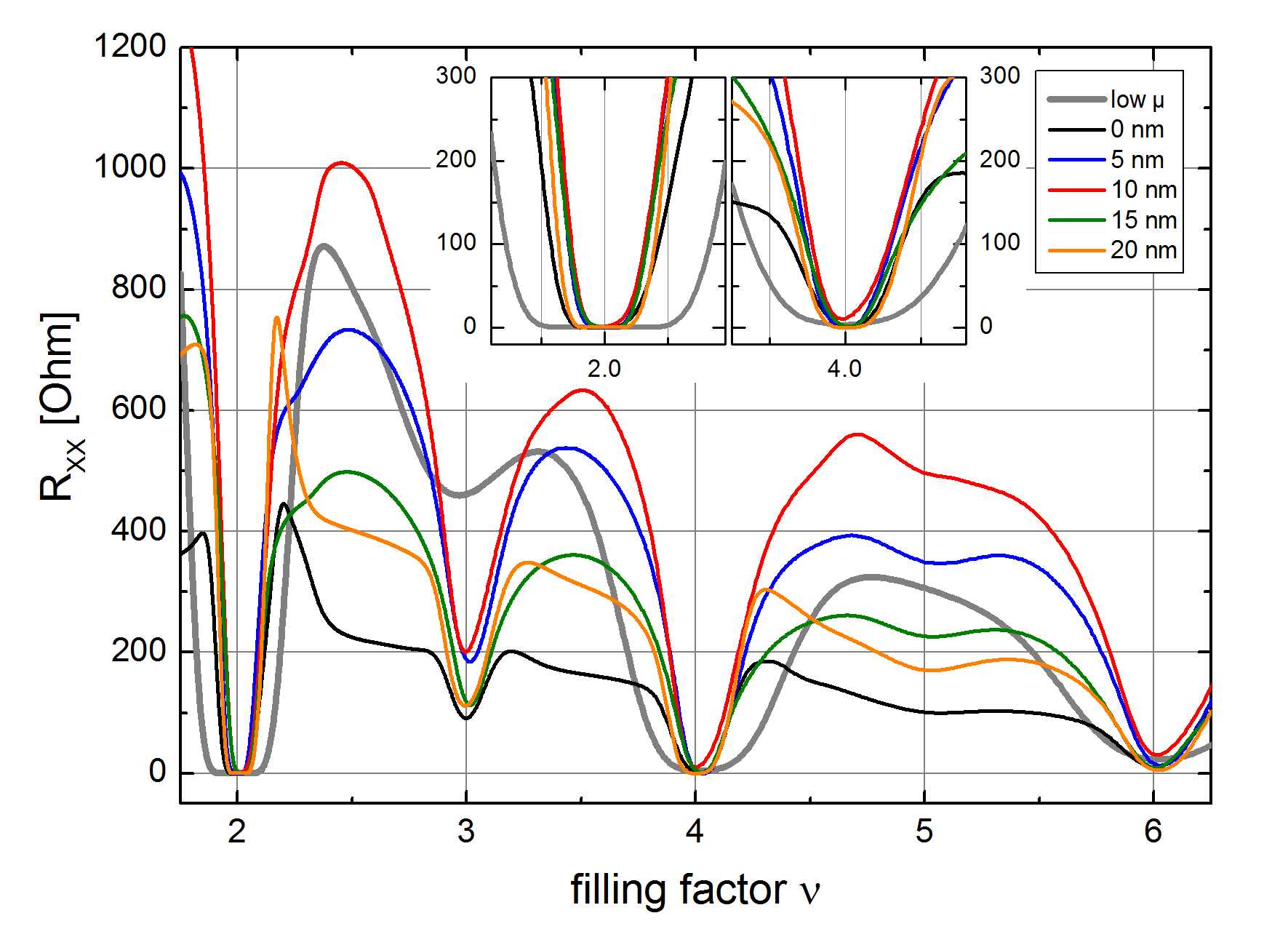} 
\par\end{centering}

\protect\protect\caption{Longitudinal resistance in the magnetic field range corresponding
to filling factor $\nu=2$ to $6$. No widening of the minima is observed
for the different positions of the Al $\delta$-layers (colour-coded).
For comparison, the grey line labelled as ``low $\mu$ '' represents
the $R_{XX}$ trace of a sample with low mobility -- very similar
to the $a=10\,\mathrm{nm}$ sample -- without any Al-doping; the minima
here are significantly broader. The increase in resistance between
the minima seems to be more related to the absolute scattering rate
rather than to the position of the Al $\delta$-layer.\label{SdHs} }
\end{figure}

Increasing the scattering rate is therefore expected to increase the
density of localized states at the expense of the extended ones and
to lead to a widening of the SdH minima in the range of the IQHE plateaus.
This behaviour is demonstrated by the trace corresponding to the sample
``low $\mu$ '' (grey line in Fig.~\ref{SdHs}) which has been
grown in an MBE system that was in a poor state at the time of growth,
i.e. which contains a high number of residual charged and neutral
impurities. It's electron mobility of $0.7\cdot10^{6}\,\mathrm{cm^{2}/Vs}$
is similar to the one of the $a=10\,\mathrm{nm}$ sample (represented
by the red line). One might expect a similar widening of minima from
samples with an Al $\delta$-layer having a comparable mobility.

We have measured the magnetotransport characteristics at $1.3\, K$
up to $6$ $Tesla$ for our samples. The resulting longitudinal resistances
as function of filling factor are shown in Fig.~\ref{SdHs}. Clearly,
no significant widening of the minima at integer filling with the
scattering rate is observed, although the scattering rates vary by
a full order of magnitude. In contrast, the maxima between the integer
filling increase considerably with the scattering rates.

The distance of the $\delta$-layer from the interface seems to be
more relevant for the shape of the curves in the regions between the
integer fillings. It would be of interest to study this behaviour
as function of (lower) temperature and compare the results with the
scaling study of Li et al.\cite{2005Li}. This is however beyond the
scope of this work. It is noteworthy that also fractional quantum
Hall effect gaps, measured by Deng et al. \cite{2014aDeng}, showed
surprisingly little change from moderate but homogeneous Al doping
which may be related to the lack of the localized-states background.

\section{Conclusions}

Placing $\delta$-layers of Al impurities into GaAs in the regions
of the 2DEG leads to substantially enhanced electron scattering rates.
The dependence of these scattering rates precisely images the shape
of the wave function $\chi(z)$, verifying that the scattering potential
acts very locally on the electron wave function. This behavior makes
this simple technique a unique way to map out the spatial distribution
of 2DEG wave functions.

Although the scattering rate due to the Al atoms was enhanced by a
factor of $10$ compared to the reference sample, it does not influence
the width of the IQHE plateaus. This indicates that this scattering
process does not contribute to the background of localized states
between the Landau levels. The Al atoms do however enhance the resistance
maxima between the integer filling factors. This indicates that the
Al atoms cause a purely elastic scattering process. The missing of
an increase of the localized background may also be relevant for the
observation by Deng et al. that neutral background impurities -- in
the form of a homogeneous Al-doping -- do not have a significant impact
on the activation energy of the $\nu=5/2$ FQHS\cite{2014aDeng}.

Using this technique it will be possible to map out wave functions
of rectangular quantum wells which are of special interest for higher
mobilities. Such structures are the testbed for investigations on
the exotic $\nu$ = 5/2 state, whose quality is currently limited
by the influence of remote ionized donors \cite{2009Umansky,2013Muraki}.
Their effect would be minimal on a symmetric wave function. Currently
such a symmetry can only be aimed at by calculating the required upper
and lower doping density, but is very difficult to verify.

Furtheron, the technique can be used for wide quantum wells and double-quantum
well systems. In such systems, the local electron density distribution
develops two maxima that need to be balanced. Again, carefully placed
Al $\delta$-layers would be helpful as a sensor to optimize the growth
parameters to achieve a balancing between two (partial) wave functions.
We acknowledge stimulating discussions with Yongqing Li, Fabian Schläpfer
and Lars Tiemann. The cooperation with Matthias Berl, Stefan Faelt,
Jessica Gmür, Siegfried Heider, Marcel Sturzenegger was essential
for operating the MBE system at the high quality level. We gratefully
acknowledge the financial support of the Swiss National Foundation
(Schweizerischer Nationalfonds, NCCR ``Quantum Science and Technology'').
This work was supported by the Academy of Finland through its Center
of Excellence program, and by the European Research Council (Grant
No. 240362-Heattronics).

 \bibliographystyle{spiebib_rev}
\bibliography{Alloy}

\end{document}